\documentclass[conference]{IEEEtran}
\IEEEoverridecommandlockouts
\usepackage{cite}
\usepackage{amsmath,amssymb,amsfonts}
\usepackage{graphicx}
\usepackage{textcomp}
\usepackage{xcolor}
\def\BibTeX{{\rm B\kern-.05em{\sc i\kern-.025em b}\kern-.08em
    T\kern-.1667em\lower.7ex\hbox{E}\kern-.125emX}}

\usepackage{cite}
\usepackage{amsmath,amssymb,amsfonts}
\usepackage{graphicx}
\usepackage{xcolor}
\usepackage{makecell}
\usepackage{booktabs} 
\usepackage{subfig}
\usepackage{graphicx}
\usepackage{amsmath}
\usepackage[normalem]{ulem}
\usepackage{enumitem}
\usepackage{multirow}
\usepackage[nomargin,inline,marginclue,draft]{fixme}
\usepackage{balance}
\usepackage{changepage}

\usepackage{balance}
\usepackage[breaklinks=true,bookmarks=false]{hyperref}
\newcommand{\para}[1]{{\vspace{1pt} \bf \noindent #1 \hspace{10pt}}}
\usepackage{multirow}
\usepackage{algorithm}
\usepackage{algpseudocode}

\usepackage{etoolbox}
\makeatletter
\patchcmd{\@makecaption}
  {\scshape}
  {}
  {}
  {}
\makeatother

\begin{document}

\title{Session-aware Item-combination Recommendation with Transformer Network}

\author{
\IEEEauthorblockN{Tzu-Heng Lin}
\IEEEauthorblockA{
\textit{Xiaomi AI Lab} \\
\textit{Xiaomi Inc.}\\
lzhbrian@gmail.com}
\and
\IEEEauthorblockN{Chen Gao}
\IEEEauthorblockA{
\textit{Department of Electronic Engineering} \\
\textit{Tsinghua University}\\
chgao96@gmail.com}
}

\IEEEpubid{978-1-6654-3902-2/21/\$31.00~\copyright~2021 IEEE}

\maketitle

\begin{abstract}
    In this paper, we detailedly describe our solution for the IEEE BigData Cup 2021: RL-based RecSys (Track 1: Item Combination Prediction)\footnote{\href{https://www.kaggle.com/c/bigdata2021-rl-recsys/}{https://www.kaggle.com/c/bigdata2021-rl-recsys/}}.
    We first conduct an exploratory data analysis on the dataset and then utilize the findings to design our framework.
    Specifically, we use a \textbf{two-headed transformer-based network} to predict user feedback and unlocked sessions, along with the proposed \textbf{session-aware reweighted loss}, \textbf{multi-tasking with click behavior prediction}, and \textbf{randomness-in-session augmentation}.
    In the final private leaderboard on Kaggle, our method ranked 2nd with a categorization accuracy of 0.39224.\footnote{Our code is available at \href{https://github.com/lzhbrian/bigdatacup2021}{https://github.com/lzhbrian/bigdatacup2021}}
\end{abstract}


\section{Introduction} \label{sec:intro}

The task of the IEEE BigData Cup 2021: RL-based RecSys (Track 1: Item Combination Prediction) \cite{2021RL4RS, kaggle} is to predict each user's purchasing feedback to nine exposed items, given this user's click history, portrait features, and items' features, which is similar to \textit{bundle recommendation}~\cite{chang2020bundle}.
The special setting in this task is that the nine items are grouped into three sessions. 
The user can only unlock the subsequent session after he/she buys all three items in the current session.

\begin{figure}[t!]
    \centering
    \includegraphics[width=\linewidth]{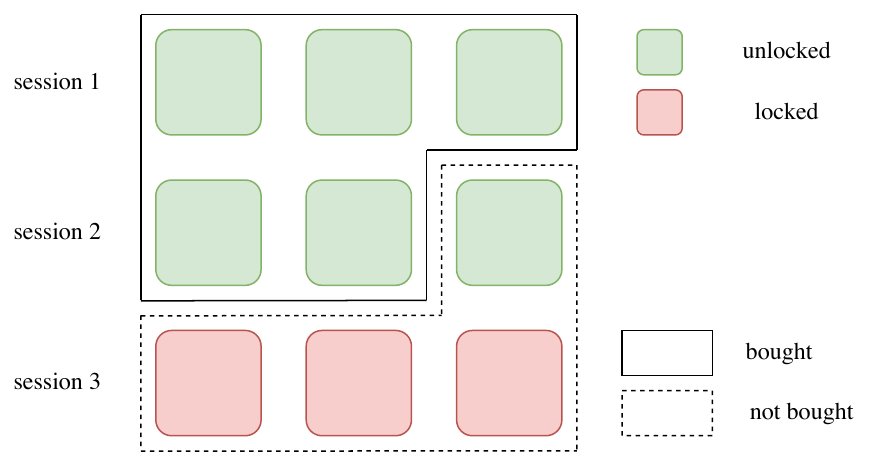}
    \caption{Problem Setup. Each user is exposed to nine items simultaneously. However, the items are divided into three 3-length sessions. The user can only unlock the subsequent three items after he/she buys all three items in the current session.
    We want to predict whether a user would buy the nine exposed items or not.}
    \label{fig:problemdef}
\end{figure}

More formally, given a user $u$ (along with his/her clicking history $c_{u,1}, c_{u,2}, ...$, and some portrait features $f_{u,1}, f_{u,2}, ..., f_{u,10}$), and his/her nine exposed items $i_{u,1}, i_{u,2}, ..., i_{u,9}$ (along with some item features $f_{i,1}, f_{i,2}, ..., f_{i,6}$ for each item $i$), 
the objective is to predict nine interactions $y_{u,1}, y_{u,2}, ..., y_{u,9} \in \{0,1\}$.
Each one of the interactions indicates whether this user would buy the corresponding item or not. 
In addition, in this scenario, the middle three items $i_{u,4}, i_{u,5}, i_{u,6}$ are not unlocked until the user has bought all of the first three items $i_{u,1}, i_{u,2}, i_{u,3}$, and similarly, the last three items $i_{u,7}, i_{u,8}, i_{u,9}$ are not unlocked until the user has bought all of the first six items $i_{u,1}, i_{u,2}, ..., i_{u,6}$ (\textit{c.f.} Figure \ref{fig:problemdef}). 
The evaluation metric for this task is the Categorization Accuracy measure, which is defined as follows,
\begin{equation}
    \textbf{accuracy} = \frac{1}{M} 
    ~\overset{M}{\underset{u=1}{\sum}}  ~\overset{9}{\underset{j=1}{\prod}} 
    ~[y_{u,j} = \hat{y}_{u,j}],
\end{equation}
where $M$ denotes the number of users, $y_{u,j}$ and $\hat{y}_{u,j}$ are the predicted and ground-truth interactions, and $[y_{u,j} = \hat{y}_{u,j}]$ is the \textit{Iverson bracket}.

Overall speaking, this task is challenging in two aspects.
\begin{itemize}
    \item Firstly, the nine exposed items are correlated and treated differently by the users. We cannot simply apply a single traditional recommendation method to predict each interaction independently.
    \item Secondly, with the given evaluation metric, it is required to correctly predict all of the nine interactions of a user, while partially correct predictions contribute nothing to the final score.
\end{itemize}

\IEEEpubidadjcol

To overcome the above challenges, we propose a delicate two-headed transformer-based framework to predict both users' buying behavior and unlocked sessions.
The unlocked session prediction can be used to refine unreasonable buy predictions.
We further propose a randomness-in-session augmentation technique and a novel session-aware reweighted loss to address the unique characteristics in this scenario.
Finally, a multi-tasking training procedure with click prediction is utilized to assist the learning of embedding layers.
Extensive experiments and ablation studies have demonstrated the effectiveness of our method.

In what follows, we will discuss related works in Section \ref{sec:relatedwork}, conduct an exploratory data analysis in Section \ref{sec:eda}, describe our proposed method in Section \ref{sec:method}, and finally conclude the paper with discussion and future works in Section \ref{sec:conclusion}.

\section{Related work} \label{sec:relatedwork}
Recommender systems aim to filter information for users, which has become one kind of fundamental service in today's information platforms~\cite{resnick1997recommender}.
Generally, from the perspective of real-world application, the recommender systems contain two stages, matching and ranking. 
Recently, deep learning has become the state-of-the-art solution of recommender systems in both two stages~\cite{dlrs,wu2021survey,gao2021graph}.
As for the matching stage, of which the mainstream methods are collaborative filtering~\cite{su2009survey}, which learns user interests from historical behaviors, deep neural networks methods~\cite{he2017neural}, or even graph neural networks~\cite{he2020lightgcn,wang2020disentangled}, achieve promising performance.
As for the ranking stage, which is also known as click-through rate (CTR) prediction, deep learning-based models such as DeepFM with multi-layer perceptron~\cite{deepfm}, xDeepFM with compressed interaction network~\cite{lian2018xdeepfm}, DIN~\cite{din} with attention mechanisms, etc., are demonstrated effective in learning from complex features of users and items.

In this work, we develop a method based on transformer network, a recent advance of neural network with extraordinary achievements in many areas, for capturing the complex behavior of users in the task of item combination recommendation.

\section{Exploratory Data Analysis} \label{sec:eda}
Before diving into the model design, we conduct exploratory data analysis firsthand to master the whole picture of the dataset.

\subsection{Data statistics}
Table \ref{tab:datastats} shows the overall statistics of this dataset. 
In total, there are 381 items.
There are 260,087 buying entries for training and 206,254 buying entries for testing. These entries are also accompanied by 10,435,798 and 8,357,719 clicking logs, respectively. We will then analyze more details about the clicking and buying behavior of users in the following.

\begin{table}[t!]
    \centering
    \caption{Overall data statistics}
    \begin{tabular}{c|c|c|c|c}
        \hline
        \multicolumn{2}{c|}{\# buying entries (users)} &
        \multicolumn{2}{c|}{\# clicks} &
        \multirow{2}{*}{\# items} \\
        \cline{1-4}
          \# train & \# test & \# train & \# test & \\
        \hline
        260,087 & 206,254 & 10,435,798 & 8,357,719 & 381 \\
        \hline
        \end{tabular}
    \label{tab:datastats}
\end{table}

Table \ref{tab:clickbuysess} shows how many clicks and buys do items in each session possess. It's worth noticing that an item would only appear in its specific session.
We can see that items in later sessions are with more types, and items with earlier sessions possess more clicks and buys. This is reasonable since users need to buy early items in order to unlock items (with higher prices) in the later sessions.

\begin{table}[t!]
    \centering
    \caption{Click and buy statistics in different sessions.}
    \begin{tabular}{c|c|c|c|c} 
    \hline
    {session} & {item IDs} & {\# items} & {\# clicks} & {\# buys} \\ \hline
    {1} & {1$\sim$39}& 39 & {4,606,977} & {616,952} \\ \hline
    {2} & {40$\sim$147} & 108 & {3,608,173} & {485,449} \\ \hline
    {3} & {148$\sim$381} & 234 & {2,220,648} & {287,482} \\ \hline
    \end{tabular}
    \label{tab:clickbuysess}
\end{table}

\subsection{Buying behavior analysis} \label{sec:eda:buyanalysis}
\begin{figure}[t!]
    \centering
    \includegraphics[width=0.8\linewidth]{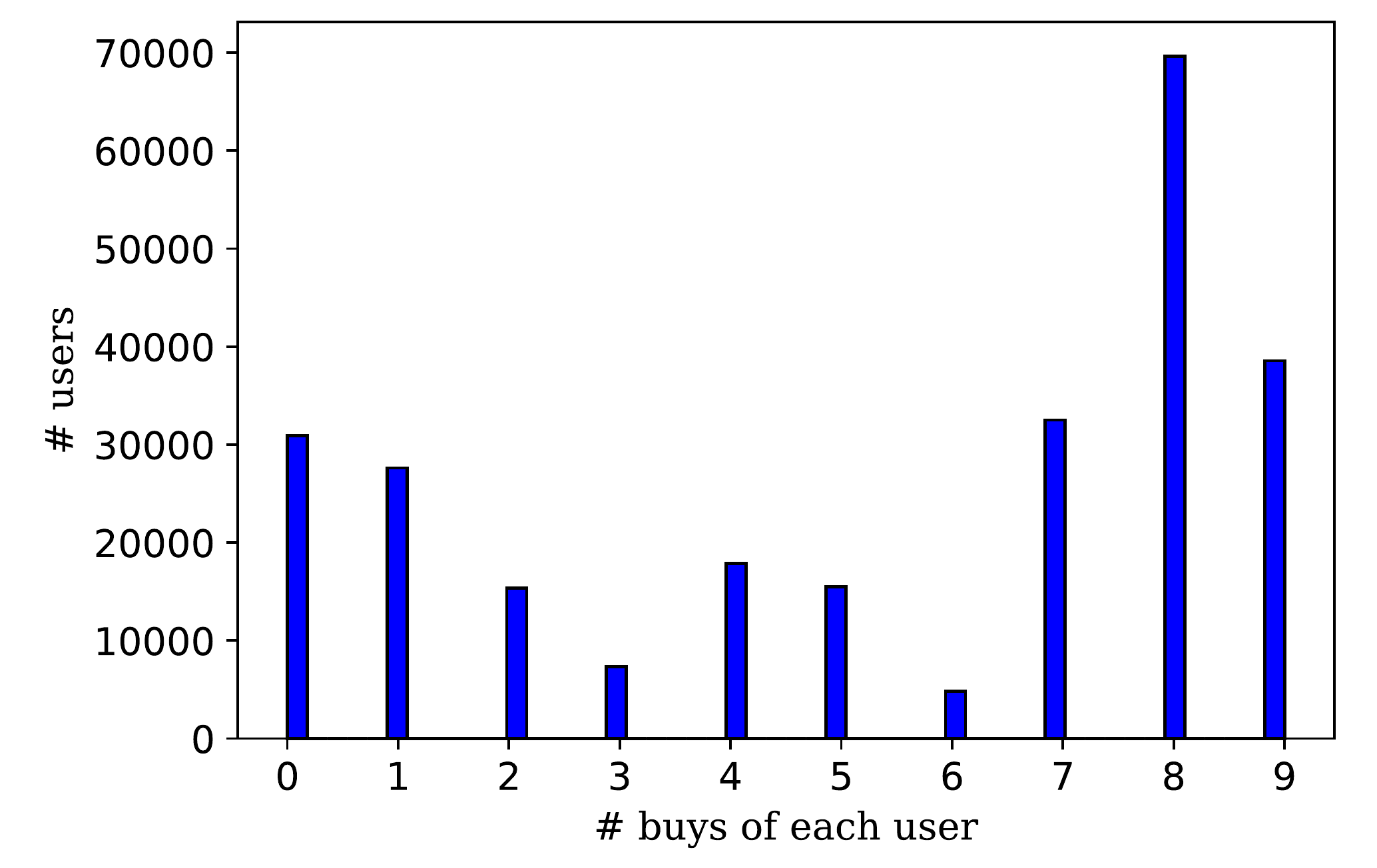}
    \caption{Histogram of the number of buys of each user.}
    \label{fig:analysis:buy}
\end{figure}

Due to the dataset characteristics (\textit{c.f.} Section \ref{sec:intro}), we plot the histogram of the number of items each user bought in Fig. \ref{fig:analysis:buy}, and classify users into four groups according to the number of items they have bought as follows,
\begin{itemize}
    \item Group-0: 30,912 users who have bought 0 item.
    \item Group-1: 50,267 users who have bought 1$\sim$3 items.
    \item Group-2: 38,191 users who have bought 4$\sim$6 items.
    \item Group-3: 140,717 users who have bought 7$\sim$9 items.
\end{itemize}
We can see that a decent population (Group-0) didn't buy anything, the number of users who bought 4$\sim$6 items (Group-2) are the fewest, and a large portion of users (Group-3) chose to buy no less than seven items. This indicates an \textbf{hourglass shape of user distribution}.
It's also worth noticing that very few people buy three or six items (\textit{c.f.} Fig. \ref{fig:analysis:buy}). We hypothesize that this is because the main reason why a user buys three or six items is to unlock and buy items in the next session.

\subsection{Clicking behavior analysis} 
\begin{figure}[t!]
    \centering
    \includegraphics[width=0.8\linewidth]{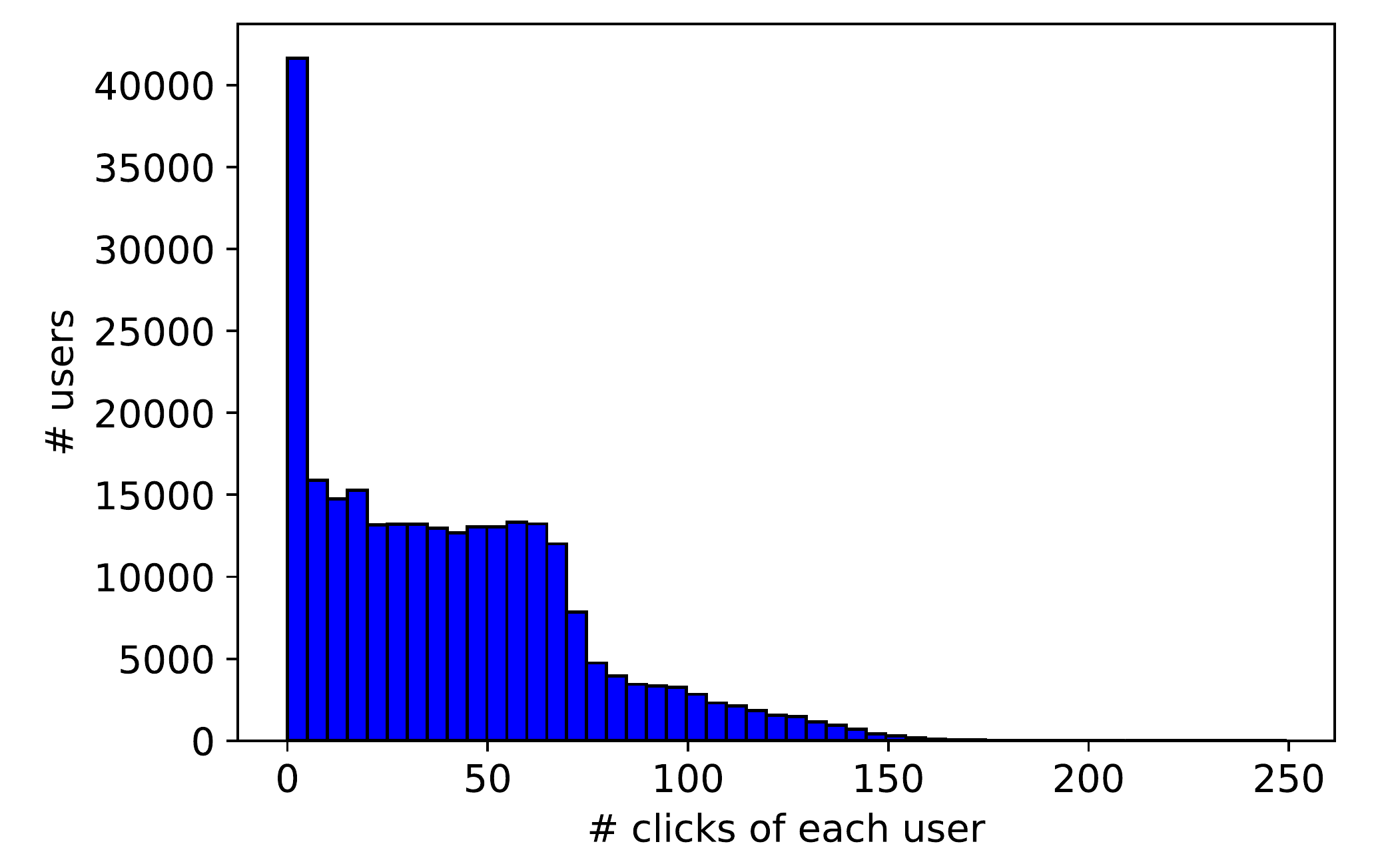}
    \caption{Histogram of the number of clicks of each user.}
    \label{fig:analysis:click}
\end{figure}

We plot the histogram of the number of clicks of each user in Fig. \ref{fig:analysis:click}.
There are 28,184 users who did not click anything. However, we do see that the majority of users are with a decent number of clicks, which motivates us to utilize the clicking logs to assist the training.

\subsection{User portrait features and item features}
We further present user portrait features and item features in Table \ref{tab:user_portrait} and Table \ref{tab:item_features}.
We can see that all user portraits are discrete features, while two of the item features are continuous features.

\begin{table*}[t!]
    \centering
    \caption{User portrait features.}
    \begin{tabular}{c|c|c|c|c|c|c|c|c|c|c}
    \hline
    {\textbf{user features}} & 
    { $f_{u,1}$} & 
    { $f_{u,2}$} & 
    { $f_{u,3}$} & 
    { $f_{u,4}$} & 
    { $f_{u,5}$} & 
    { $f_{u,6}$} & 
    { $f_{u,7}$} & 
    { $f_{u,8}$} & 
    { $f_{u,9}$} & 
    { $f_{u,10}$} \\ \hline
    { \# unique values in train set} & 3 & { 1363} & { 20} & { 10} & { 195} & { 49} & { 3} & { 11} & { 2} & { 2164} \\ \hline
    { \# unique values in test set} & 3 & { 1319} & { 19} & { 10} & { 191} & { 47} & { 3} & { 13} & { 2} & { 2054} \\ \hline
    discrete or continuous (disc./cont.) & disc. & disc. & disc. & disc. & disc. & disc. & disc. & disc. & disc. & disc. \\ \hline
    \end{tabular}
    \label{tab:user_portrait}
\end{table*}

\begin{table*}[t!]
    \centering
    \caption{Item features.}
    \begin{tabular}{c|c|c|c|c|c|c}
    \hline
    { \textbf{item features}} & { $f_{i,1}$} & 
    { $f_{i,2}$} & 
    { $f_{i,3}$} & 
    { $f_{i,4}$} & 
    { $f_{i,5}$} & 
    { $f_{i,6}$ (price)} \\ \hline
    {\# unique values} & {4} & {10} & {2} & {n/a} & {n/a} & {248} \\ \hline
    {values} & {1,2,3,4} & {0,1,2,3,4,5,6,7,8,9} & {1,2} & { 0$\sim$1, float} & {0$\sim$1, float} & {150$\sim$16621, int} \\ \hline
    discrete or continuous (disc./cont.) & disc. & disc. & disc. & cont. & cont. & cont. \\ \hline
    \end{tabular}
    \label{tab:item_features}
\end{table*}

\section{Method \& Experiments} \label{sec:method}

\begin{table*}[t!]
    \centering
    \caption{Experimental results of different models (take G as an example, it is built on F with an additional design of augmentation). The numbers in this table are ablation studies after the competition. * means the settings of the best submission during competition. $\diamond$ means the settings are providing unstable yet higher scores.}
    \begin{tabular}{l|c|c} \hline
        Model & validation & test \\\hline
        \textbf{A}~~MLP basic model & 0.29169 & 0.33817 \\
        \textbf{B}~~+ randomness-in-session augmentation (train) & 0.29965 & 0.35007 \\
        \textbf{C}~~+ transformer backbone & 0.31140 & 0.36210 \\
        \textbf{D}~~+ two-headed (buy and group) prediction & 0.31475 & 0.36258 \\
        \textbf{E}~~+ session-aware loss reweighting & 0.33090 & 0.38355 \\
        \textbf{F}~~+ multi-tasking with click prediction * & 0.33323 & 0.38805 \\
        \textbf{G}~~+ randomness-in-session augmentation (inference) $\diamond$ & 0.33335 & 0.39161 \\\hline
    \end{tabular}
    \label{tab:ablation}
\end{table*}

\begin{figure*}[htbp]
    \centering
    \includegraphics[width=\linewidth]{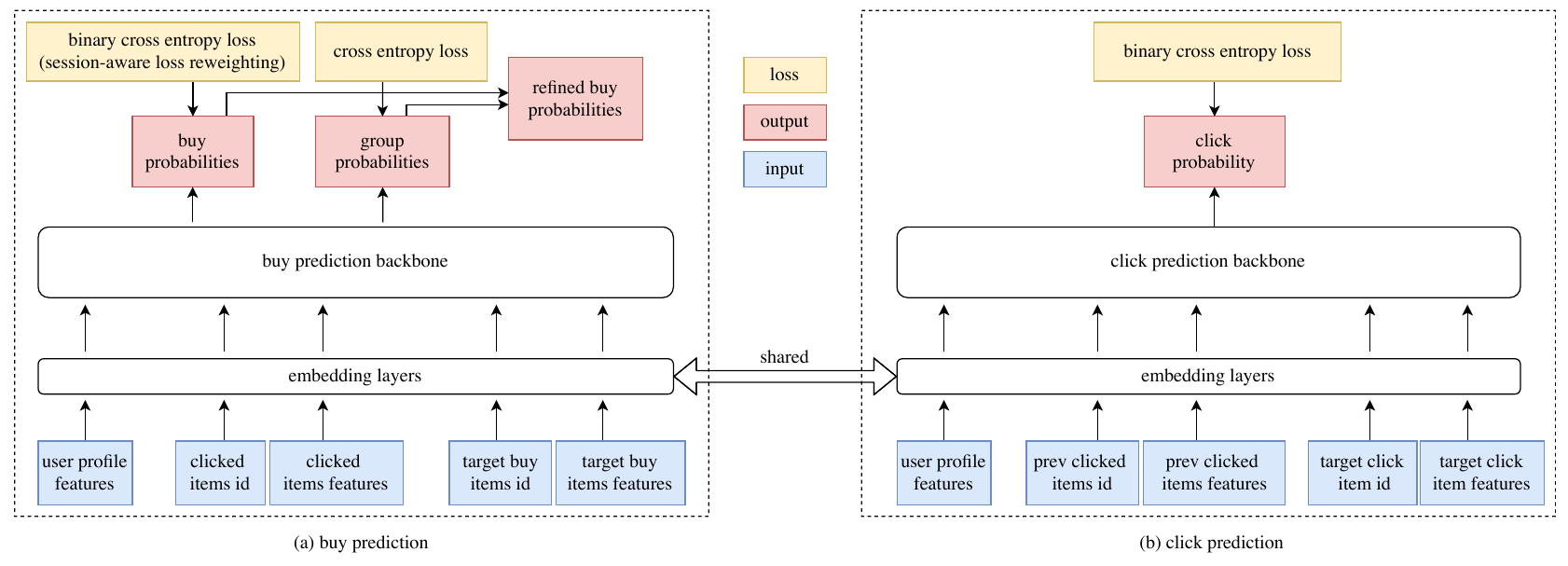}
    \caption{Overall structure of our method.}
    \label{fig:net}
\end{figure*}

The overall structure of our method is shown in Fig. \ref{fig:net}. 
In what follows, we will introduce each part of our framework.
The ablation study results are shown in Table \ref{tab:ablation}.

\subsection{Network design}

\para{MLP basic model (\textbf{Config-A})} 
We start with a very simple basic network.
The network takes the following inputs: user profile features, user clicked items' id and features, nine exposed target items' id and features.
These inputs are processed by their corresponding embedding layers, and further fed to an MLP module.
Then the network predicts whether the user will buy the nine exposed target items.
We propose this framework since the nine items' labels are correlated. For example, users might buy all of the first six items, only to unlock and buy subsequent items. Therefore, it is not suitable to predict the nine feedback independently, and we need to ensure the network is able to predict nine feedback simultaneously.
The training of the model is supervised by a vanilla binary cross entropy (BCE) loss on each item respectively as follows,
\begin{equation}
    \mathcal{L}_{\text{buy}} = \frac{1}{M} ~\overset{M}{\underset{u=1}{\sum}}
    ~\overset{9}{\underset{j=1}{\sum}} 
        ~\text{BCE}(\hat{y}_{u,j}, y_{u,j}),
\end{equation}
where $\hat{y}_{u,j}$ and $ y_{u,j}$ denote the ground-truth and predicted feedback between user $u$ and the exposed $j$-th item, respectively, and
\begin{equation}
\begin{aligned}
    \text{BCE}(\hat{y}_{u,j}, y_{u,j}) = 
        &- \hat{y}_{u,j} \log y_{u,j} \\
        &- (1-\hat{y}_{u,j}) \log (1-y_{u,j})
\end{aligned}
\end{equation}
is the binary cross entropy term for each one of the nine items.
We set the embedding size to 16 here, and the MLP-structure is set to $\{$1440, 256, 64, 9$\}$. This very simple basic model can achieve 0.29169 on validation set, and 0.33817 on test set.

\para{Randomness-in-session augmentation (\textbf{Config-B}, \textbf{G})}
To prevent over-fitting and make training more robust, we randomly shuffle items' orders within the same sessions during the training. 
Note that in this scenario, users are not sensitive to the items' order within the same session. However, our network treats them with different parameters. So we propose to use this augmentation technique to alleviate this shortcoming.
This strategy is also used for test time augmentation, where original prediction and predictions produced by shuffled inputs are averaged to produce the final results.
In our experiments, augmentation in training (\textbf{Config-B}) increases the score from 0.29169 to 0.29965 on validation set and from 0.33817 to 0.35007 on test set.
However, this proposed augmentation method in inference is not so stable and sometimes might do some harm to the score.
In Table \ref{tab:ablation}, we are just reporting the result of one experiment (\textbf{Config-G}), which improves the score.

\para{Transformer backbone (\textbf{Config-C})}
Instead of simple MLPs \cite{din, ncf}, we switch the backbone part into a transformer \cite{transformer}, as their self-attention mechanism is proved to be effective on capturing inter-relations between different features.
\begin{equation}
\begin{aligned}
\mathbf{z}_{0} &=\left[x^{1} \mathbf{E} ; x^{2} \mathbf{E}; \cdots; x^{N} \mathbf{E}\right]+\mathbf{E}_{\text{pos}}, &  \\
\mathbf{z}_{\ell} &=\operatorname{MSA}\left(\operatorname{LayerNorm}\left(\mathbf{z}_{\ell-1}\right)\right)+\mathbf{z}_{\ell-1}, & \ell=1 \ldots L \\
\mathbf{z}_{\ell} &=\operatorname{MLP}\left(\operatorname{LayerNorm}\left(\mathbf{z}_{\ell}\right)\right)+\mathbf{z}_{\ell}, & \ell=1 \ldots L \\
\mathbf{y} &=\operatorname{LayerNorm}\left(\mathbf{z}_{L}\right), &
\end{aligned}
\end{equation}
where $x^{n}$ is the corresponding one-hot vector of features, $\mathbf{E} \in \mathbb{R}^{N \times D}, \mathbf{E}_{pos} \in \mathbb{R}^{N \times D}$, $N$ is the number of features, $D$ is the embedding size, $L$ is the number of transformer layers, and $\text{MSA}(\cdot)$ is the multi-head self attention.
We set the embedding size to 128, number of layers to 3, number of self-attention head to 4, and the sizes of $q$,$k$,$v$ in the self-attention module to 32, and MLP size to 64.
Using the transformer backbone (\textbf{Config-C}) can improve our score from 0.29965 to 0.31140, and from 0.35007 to 0.36210 on validation and test set, respectively.
However, we do note that this improvement compared to \textbf{Config-B} might in part come from a larger embedding and network size. We didn't do that ablation study due to limited time.

\para{Two-headed (buy and group) prediction (\textbf{Config-D})}
One should notice that the above simple framework might introduce some invalid buy predictions that are impossible to happen in the real world.
For example, the network might predict that the user buys two items, the 1st one and the 9th one. 
However, this is impossible since the user has to buy all of the first 6 items in order to buy the 9th item.

Thus, in addition to the buy prediction, we propose to also predict the group (as defined in Section \ref{sec:eda:buyanalysis}) of each user, which forms a two-headed prediction network, as shown in Fig. \ref{fig:net}(a).
This group prediction part is supervised by a cross entropy loss as follows,
\begin{equation}
    \mathcal{L}_{\text{group}} = \frac{1}{M} ~\overset{M}{\underset{u=1}{\sum}}
    ~\overset{4}{\underset{j=1}{\sum}}
    - \hat{g}_{u,j} \log g_{u,j},
\end{equation}
where $\mathbf{\hat{g}}_{u} \in \mathbb{R}^4$ is a one-hot ground-truth vector indicating which group user $u$ belongs to. Here $\mathbf{g}_u \in \mathbb{R}^4$ is the predicted group vector (after a \textit{softmax} layer).
The loss is added with previous ones and back-propagated together as follows,
\begin{equation}
    \mathcal{L} = 
    \lambda_{\text{buy}} \mathcal{L}_{\text{buy}} + 
    \lambda_{\text{group}} \mathcal{L}_{\text{group}},
\end{equation}
where we set $\lambda_{\text{buy}}=0.8, \lambda_{\text{group}}=0.1$. 
After training, the predicted group vector $\mathbf{g}_u$ will be used to refine and fix the unreasonable predicted buying behavior of the nine exposed items $\mathbf{y}_{u} \in \mathbb{R}^{9}$ as follows,
\small
\begin{equation}\label{eqn:refine}
    \mathbf{y}_{u} = 
    \begin{cases}
        [0,0,0, 0,0,0, 0,0,0] & \arg\underset{j}{\max}~\mathbf{g}_{u}=0\\
        [y_{u,1},y_{u,2},y_{u,3}, 0,0,0, 0,0,0] & \arg\underset{j}{\max}~\mathbf{g}_{u}=1\\
        [1,1,1, y_{u,4},y_{u,5},y_{u,6}, y_{u,7},y_{u,8},y_{u,9}] & \arg\underset{j}{\max}~\mathbf{g}_{u}=2\\
        [1,1,1, 1,1,1, y_{u,7},y_{u,8},y_{u,9}] & \arg\underset{j}{\max}~\mathbf{g}_{u}=3.
    \end{cases}
\end{equation}
\normalsize
After refined using the group predictions, our score improves from 0.31140 to 0.31475 on validation set and from 0.36210 to 0.36258 and test set.

\begin{figure}[!t]
    \centering
    \includegraphics[width=\linewidth]{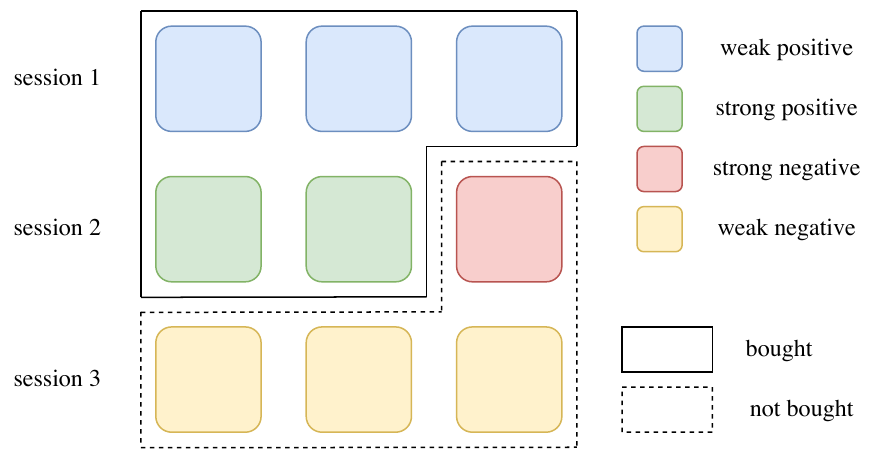}
    \caption{Session-aware loss reweighting}
    \label{fig:reweightloss}
\end{figure}

\para{Session-aware loss reweighting (\textbf{Config-E})}
To better model users' buying behaviors, we classify the nine exposed items into four types (weak positive, strong positive, strong negative, weak negative) as shown in Fig. \ref{fig:reweightloss}.
\begin{itemize}
    \item For sessions before the last session user has unlocked, items should be treated as \textbf{weak positives}, as the user might buy these items only to unlock the later sessions.
    \item For the last session user has unlocked, items should be treated as \textbf{strong positives} and \textbf{strong negatives}. As the user unlocked and stopped in this session, items bought or not bought should be classified as strong signals.
    \item For later locked sessions, items should be treated as \textbf{weak negatives}, as users haven't unlocked these sessions, we should not assume too strong preferences on these items.
\end{itemize}
In practice, we assign different weights $\lambda_1, \lambda_2, \lambda_3, \lambda_4$ for the above 4 types of items. The formally defined loss can be written as follows,
\small
\begin{equation}
\begin{aligned}
\centering
 &\quad \quad \quad \quad \quad  \quad \quad \quad     \mathcal{L}_{\text{buy-reweight}} = \frac{1}{M} ~\overset{M}{\underset{u=1}{\sum}} \\
        &\begin{cases}
        ~\overset{9}{\underset{j=1}{\sum}} ~\lambda_4 \Gamma_{u,j}
        & \mathbf{\hat{g}}_{u} = [1,0,0,0] \\
        ~\overset{3}{\underset{j=1}{\sum}} ~\Lambda_{\lambda_2,\lambda_3,u,j} + 
        ~\overset{9}{\underset{j=4}{\sum}} ~\lambda_4 \Gamma_{u,j}
        & \mathbf{\hat{g}}_{u} = [0,1,0,0] \\
        ~\overset{3}{\underset{j=1}{\sum}} ~\lambda_1 \Gamma_{u,j} + 
        ~\overset{6}{\underset{j=4}{\sum}} ~\Lambda_{\lambda_2,\lambda_3,u,j} + 
        ~\overset{9}{\underset{j=7}{\sum}} ~\lambda_4 \Gamma_{u,j}
        & \mathbf{\hat{g}}_{u} = [0,0,1,0] \\
        ~\overset{6}{\underset{j=1}{\sum}} ~\lambda_1 \Gamma_{u,j} + 
        ~\overset{9}{\underset{j=7}{\sum}} ~\Lambda_{\lambda_2,\lambda_3,u,j}
        & \mathbf{\hat{g}}_{u} = [0,0,0,1] \\
        \end{cases}, \\
\end{aligned}
\end{equation}
\normalsize
where $\Gamma_{u,j}$ and $\Lambda_{\lambda_2,\lambda_3,u,j}$ denote losses for weak positive/negative items and strong positive/negative items, respectively, which are formulated as follows,
\begin{equation}
\footnotesize
\begin{aligned}
  \Gamma_{u,j} =& ~\text{BCE}(\hat{y}_{u,j}, y_{u,j})\\
  \Lambda_{\lambda_2,\lambda_3,u,j} =&  ~\lambda_2~\hat{y}_{u,j}~\text{BCE}(\hat{y}_{u,j}, y_{u,j})+ \\
  &~\lambda_3~(1-\hat{y}_{u,j})~\text{BCE}(\hat{y}_{u,j}, y_{u,j}).
\end{aligned}
\end{equation}
In our experiments, we replace the original $\mathcal{L}_{\text{buy}}$ with $\mathcal{L}_{\text{buy-reweight}}$, and set $\lambda_1 = 0.5, \lambda_2 = 1, \lambda_3 = 1, \lambda_4 = 0.5$.
This design can greatly boost our score from 0.31475 to 0.33090 on validation set, and from 0.36258 to 0.38355 on test set.

\para{Multi-tasking with click prediction (\textbf{Config-F})}
Apart from the buy prediction network described in Fig. \ref{fig:net}(a), we propose to use another click prediction auxiliary network (Fig. \ref{fig:net}(b)) to assist the learning procedure.
Note that the two networks share the same embedding layers.
The click prediction network takes the following inputs: user profile features, the previously clicked items' id and features, target items' id, and features. It is trained to predict whether the user will click the target item or not.
The loss function is defined as follows,
\begin{equation}
    \mathcal{L}_{\text{click}} = \frac{1}{M} ~\overset{M}{\underset{u=1}{\sum}}
        ~\text{BCE}(\hat{c}_{u}, c_{u}),
\end{equation}
where $\hat{c}_{u}, c_{u}$ are groundtruth and predicted feedback from user $u$ to his/her target item, and
\begin{equation}
\begin{aligned}
    \text{BCE}(\hat{c}_{u}, c_{u}) = 
        &- \hat{c}_{u} \log c_{u}\\
        &- (1-\hat{c}_{u}) \log (1-c_{u}).
\end{aligned}
\end{equation}
is the binary cross entropy term.
The loss is added with previous ones and back-propagated together as follows,
\begin{equation}
    \mathcal{L} = 
    \lambda_{\text{buy}} \mathcal{L}_{\text{buy-reweight}} + 
    \lambda_{\text{group}} \mathcal{L}_{\text{group}} + 
    \lambda_{\text{click}} \mathcal{L}_{\text{click}},
\end{equation}
where we set $\lambda_{\text{buy}}=0.8, \lambda_{\text{group}}=0.1, \lambda_{\text{click}}=0.1$, and use the same network hyper-parameters as the buy prediction network here.
With the auxiliary click prediction network multi-tasking, our score is improved from 0.33090 to 0.33323 and 0.38355 to 0.38805 on the validation set and test set, respectively.

\para{Final submission} 
Our final best submission during the competition (0.33687 on validation set, 0.39224 on test set) is achieved by \textbf{Config-F}, as shown in in Table \ref{tab:ablation}.
That training instance shows much better performance than our ablation studies conducted after the competition. 
However, these methods are still suffering from the performance variances with different random seeds, which may be caused by the scale of the dataset.
We leave the efforts to address the issue of unstable performances as future work.

\subsection{Train/validation split by user portrait}
Although the competition guidelines want us to recognize each buying entry as an individual user, we notice that there are entries with identical clicking histories and user portrait features (which means the same user produces two entries).
Thus, it is more proper to split train and validation sets while taking the above observation into consideration.
We propose to view all entries with identical user portrait features as the same user and use 85\% users as train set and the rest 15\% users as the validation set.
This results in 243,775 and 16,312 entries for the train set and validation set, respectively.

\subsection{Other settings}
We use Adam \cite{adam} with default hyper-parameters in PyTorch \cite {pytorch}.
The batch size is set to 32, and the learning rate is set to 1e-2 for ten epochs.
Colab with one P100 GPU is used as our training platform, and each model takes about 2$\sim$3 hours to train.
Clicking data in the test set of both track-1 and track-2 are used during our training. 
Checkpoint with the best score on the validation set is used for evaluation.
All continuous features are discretized into bins.

\section{Conclusion} \label{sec:conclusion}

In this paper, we propose a framework for item combination prediction. Specifically, we propose several delicate designs to improve the performance, namely randomness-in-session augmentation, transformer backbone, two-headed prediction, session-aware loss reweighting, and multi-tasking with click prediction.
Extensive experiments have proved the effectiveness of our framework.

We have also tried several things that conceptually make sense but did not improve the score.
Firstly, we tried an attention-like deep interest network \cite{din} to reweight user clicked items, however, it didn't improve the final score.
Given that we do not know how click data is collected, we think that users might present different preferences in the scenario where click data is collected. And thus, making the model more complex in this aspect doesn't help.
Secondly, we tried to add user embedding into the network yet encountered severe over-fitting in training. Adding mini-batch aware regularization \cite{din} can reduce over-fitting, however, it still cannot make improvements to the final score.
Due to the fact that most users only have one training entry, this result is not very surprising.
In addition, we tried adding timestamp as a feature, however, it also didn't help. We originally thought that weekends or holidays might affect user behaviors.

Future works shall include in-depth analysis and utilization with the actual meaning of user features, item features, and clicking data. It would also be interesting to investigate other network architectures that could address the multi-feedback item combination prediction scenario.
Since our work does not introduce the model ensemble technique, it is also a promising direction for future works.

\balance
\bibliographystyle{IEEEtran}
\bibliography{ref}

\end{document}